\documentclass[conference,a4paper]{IEEEtran}
\IEEEoverridecommandlockouts
\usepackage{cite}
\usepackage{amsmath,amssymb,amsfonts}
\usepackage{algorithmic}
\usepackage{graphicx}
\usepackage{textcomp}
\usepackage{xcolor}

\DeclareMathOperator\sign{sign}
\DeclareMathOperator*{\argmax}{arg\,max}
\DeclareMathOperator*{\argmin}{arg\,min}
\usepackage{mathtools}
\usepackage{booktabs}
\usepackage{multirow}
\usepackage[per-mode=symbol]{siunitx}
\usepackage[acronym,smallcaps]{glossaries}
\newacronym{AED}{AED}{Automorphism Ensemble Decoding}
\newacronym{AWGN}{AWGN}{Additive White Gaussian Noise}
\newacronym{ASIC}{ASIC}{Application-Specific Integrated Circuit}
\newacronym{BER}{BER}{Bit Error Rate}
\newacronym{BP}{BP}{Belief Propagation}
\newacronym{BPSK}{BPSK}{Binary Phase Shift Keying}
\newacronym{BSMC}{BSMC}{Binary Symmetric Memoryless Channel}
\newacronym{CRC}{CRC}{Cyclic Redundancy Check}
\newacronym{DRAM}{DRAM}{Dynamic Random Access Memory}
\newacronym{FD-SOI}{FD-SOI}{Fully Depleted Silicon on Insulator}
\newacronym{FEC}{FEC}{Forward Error Correction}
\newacronym{FER}{FER}{Frame Error Rate}
\newacronym{GA}{GA}{Gaussian Approximation}
\newacronym{GRAND}{GRAND}{Guessing Random Additive Noise Decoding}
\newacronym{GSM}{GSM}{Global System for Mobile Communications}
\newacronym{IP}{IP}{Intellectual Property}
\newacronym{IoT}{IoT}{Internet of Things}
\newacronym{LDPC}{LDPC}{Low Density Parity Check}
\newacronym{LLR}{LLR}{Log-Likelihood Ratio}
\newacronym{LTE-A}{LTE-A}{Long Term Evolution Advanced}
\newacronym{MAP}{MAP}{Maximum a Posteriori}
\newacronym{ML}{ML}{Maximum Likelihood}
\newacronym{MS}{MS}{Min-Sum}
\glssetnoexpandfield{first}
\glssetnoexpandfield{firstpl}
\newglossaryentry{ORBGRAND}{
    type={acronym}, 
    sort={Ordered Reliability Bits Guessing Random Additive Noise Decoding},  
    name={ORBGRAND}, 
    short={ORBGRAND}, 
    long={Ordered Reliability Bits Guessing Random Additive Noise Decoding},
    first={
        \ifglsused{GRAND}{Ordered Reliability Bits GRAND  (ORBGRAND)}
            {\glsunset{GRAND}Ordered Reliability Bits Guessing Random Additive Noise Decoding (ORBGRAND)}}, 
    description={Ordered Reliability Bits Guessing Random Additive Noise Decoding}
}
\newacronym{PFT}{PFT}{Polar Factor Tree}
\newacronym{PVT}{PVT}{Process, Voltage and Temperature}
\newacronym{QC}{QC}{quasi-cyclic}
\newacronym{RM}{RM}{Reed-Muller}
\newacronym{SCAN}{SCAN}{Soft Cancellation}
\newacronym{SCL}{SCL}{Successive Cancellation List}
\newacronym{SC}{SC}{Successive Cancellation}
\newacronym{SNR}{SNR}{Signal-to-Noise-Ratio}
\newacronym{SOA}{SOA}{State-of-the-Art}
\newacronym{SRAM}{SRAM}{Static Random Access Memory}
\newacronym{SSC}{SSC}{Simplified SC}
\newacronym{SSCL}{SSCL}{Simplified SCL}
\newacronym{SoC}{SoC}{System-on-Chip}
\newacronym{TDP}{TDP}{Thermal design power}
\newacronym{UMTS}{UMTS}{Universal Mobile Telecommunications System}
\newacronym{URLLC}{URLLC}{Ultra-Reliable Low Latency Communications}
\newacronym{PM}{PM}{Path Metric}
\newacronym{R0}{R0}{Rate-0}
\newacronym{REP}{REP}{Repetition}
\newacronym{SPC}{SPC}{Single Parity-Check}
\newacronym{HDD}{HDD}{Hard Decision Decoding}
\newacronym{BLTA}{BLTA}{Block Lower-Triangular Affine}
\newacronym{EC}{EC}{Equivalence Class}

\DeclareSIUnit[]{\bit}{bit}

\newcommand\Gbps[1]{\SI{#1}{\giga\bit\per\second}}
\newcommand{\ie}{i.\,e.,\ }
\newcommand{\eg}{e.\,g.,\ }
\newcommand{\x}{\times}

\newcommand\nm[1]{\SI{#1}{\nano\meter}}

\newcommand{\AED}{\gls{AED}\ }
\newcommand{\CRC}{\gls{CRC}\ }

\newcommand{\LLRs}{\glspl{LLR}\ }
\newcommand{\LLR}{\gls{LLR}\ }
\newcommand{\ML}{\gls{ML}\ }
\newcommand{\PFT}{\gls{PFT}\ }

\newcommand{\PM}{\gls{PM}\ }
\newcommand{\REP}{\gls{REP}\ }
\newcommand{\RFr}{\gls{R0}\ }

\newcommand{\SCL}{\gls{SCL}\ }
\newcommand{\SC}{\gls{SC}\ }
\newcommand{\SNR}{\gls{SNR}\ }
\newcommand{\SPC}{\gls{SPC}\ }

\begin{document}

\title{
    Automorphism Ensemble Polar Code\\ Decoders for 6G URLLC
    \thanks{
        The authors acknowledge the financial support by the Federal Ministry of
        Education and Research of Germany in the project “Open6GHub” (grant
        number: 16KISK019 and 16KISK004).
    }
}

\author{
    \IEEEauthorblockN{
        Claus Kestel\IEEEauthorrefmark{1},
        Marvin Geiselhart\IEEEauthorrefmark{2},
        Lucas Johannsen\IEEEauthorrefmark{1}\IEEEauthorrefmark{3},\\
        Stephan ten Brink\IEEEauthorrefmark{2} and
        Norbert Wehn\IEEEauthorrefmark{1}
        \\
    }
    \IEEEauthorblockA{
        \IEEEauthorrefmark{1}
        University of Kaiserslautern, 67663 Kaiserslautern, Germany \\
        Email: \{kestel, wehn\}@eit.uni-kl.de
    }
    \IEEEauthorblockA{
        \IEEEauthorrefmark{2}
        Institute of Telecommunications, University of Stuttgart, 70569 Stuttgart, Germany\\
        Email: \{geiselhart,tenbrink\}@inue.uni-stuttgart.de
    }
    \IEEEauthorblockA{
        \IEEEauthorrefmark{3}
        Koblenz University of Applied Sciences, 56075 Koblenz, Germany \\
        Email: \{johannsen\}@hs-koblenz.de
    }
    \\
}

\maketitle

\begin{abstract}
    The URLLC scenario in the upcoming 6G standard requires low latency and
      ultra reliable transmission, i.e., error correction towards ML performance.
      Achieving near-ML performance is very challenging especially for short
      block lengths. Polar codes are a promising candidate and already part of
      the 5G standard. The Successive Cancellation List (SCL) decoding algorithm
      provides very good error correction performance but at the cost of high
      computational decoding complexity resulting in large latency and low area
      and energy efficiency. Recently, Automorphism Ensemble Decoding (AED) gained
      a lot of attention to improve the error correction capability. In contrast
      to SCL, AED performs several low-complexity (e.g., SC) decoding in parallel.
      However, it is an open question whether AED can compete with
      sophisticated SCL decoders, especially from an implementation perspective
      in state of the art silicon technologies. In this paper we present an
      elaborated AED architecture that uses an advanced path metric based candidate
      selection to reduce the implementation complexity and compare it to
      state of the art SCL decoders in a 12nm FinFET technology. Our AED
      implementation outperform state of the art SCL decoders by up to 4.4x in
      latency, 8.9x in area efficiency and 4.6x in energy efficiency, while
      providing the same or even better error correction performance.
\end{abstract}


\section{Introduction}

The \gls{URLLC} scenario will be one of the key enablers for 6G. Compared
to 5G, it requires lower latency and particular improved error correction
 \cite{sadmeh2020} towards \ML performance. Short block lengths enable
low latency, but typically at the cost of degraded error correction capability
\cite{yuemil2022}. Hence, one of the key challenges is to find codes, decoding
algorithms and efficient decoder architectures that enable low latency, high
energy efficiency and, at the same time, error correction performance
towards \ML performance.

Polar codes are promising candidates for this application. They are the first
class of error correction codes proven to achieve the channel capacity for
\glspl{BSMC} \cite{ari_2009} and have been adopted for the control channel in
the enhanced mobile broadband (eMBB) use case of 5G. Polar codes can be decoded
with the \SC algorithm that is a low-complexity decoding algorithm, but suffers
on error correction performance, especially for short block lengths. Hence, more
sophisticated decoding algorithms were developed. The
\SCL decoding algorithm \cite{talvar_2015} considers several decoding paths in parallel
and selects the best one according to a path metric. With a sufficient large
list size $L$, \SCL approaches \ML performance, but at the cost of a very high computational
complexity.

Recently, the \gls{GRAND} \cite{dufli_2018} algorithms got a lot of
attention. In contrast to the algorithms mentioned above, \gls{GRAND} is a
universal decoding algorithm for any code. The basic idea is to guess noise
patterns to find the correct code word. \gls{GRAND}, in principle, can achieve
\ML performance if the number of guesses is large enough. But since the number
of guesses is unknown in advance, latency and throughput can vary largely
between worst case and average. \cite{condo_2022} limits this variance, but at
the cost of a decreased error correction performance. Moreover, it was shown
that the \gls{GRAND} algorithm and its derivatives are only practical for high
\SNR regions and high code rates.

\AED is another new decoding approach that was recently published and can be
applied to any code that has automorphisms. The automorphisms are used to
permute code words. The different code words can then be decoded in parallel by
low-complexity decoding algorithms. While there exist many automorphisms for
\gls{RM} codes \cite{geielk_2021}, it is challenging to find useful
automorphisms for polar codes. Recently \cite{geielk_2021a} have shown that
automorphisms can also be found for polar codes and that the error-correction
performance can be improved significantly.

\AED is also very interesting from an implementation point of view. Instead of
one complex decoders (\eg SCL), several low-complexity decoders (\eg SC) can be
used. Each of the low-complexity decoders has a very high implementation
efficiency in terms of latency, high throughput, large energy and area
efficiency. Since the decoders run in parallel without information exchange,
locality is significantly improved. This is very beneficial for a large implementation
efficiency and, as a result, area and energy efficiency scale almost linear with
the number of ensemble decoder instances $M$, where $M$ is defined by the
requirement on the error correction performance. It is important to mention
that the latency in \AED is first order independent of $M$ that is
unique for advanced decoding.

This work makes the following new contributions:
\begin{itemize}
    \item A simple data-driven permutation selection method for \gls{AED}.
    \item An efficient \AED algorithm for very high throughput and low
        latency that is based on a Fast \gls{SSC} algorithm with a new path metric
        based candidate selection.
    \item A detailed comparison of our \AED architecture and state-of-the-art \SCL
        decoders with respect to error correction performance and hardware
        implementation on a \nm{12} FinFET silicon technology. This is, to the best
        of our knowledge, the first implementation and comparison of \AED with
        traditional decoding algorithms.
\end{itemize}

The remainder of this paper is structured as follows: We provide the required
background of Polar codes and their decoding algorithms in
Section~\ref{sec:background}. The concept of \AED and a new path metric based
candidate selection is provided in Section~\ref{sec:aed}. We present a detailed
comparison with \SCL decoders regarding error correction performance and
implementation costs in Section~\ref{sec:results} and conclude this paper in
Section~\ref{sec:conclusion}.

\section{Background}
\label{sec:background}

\subsection{Polar Codes}

Polar codes $\mathcal{P}(N,K)$ are linear block codes with code length $N=2^n$,
which encode $K$ information bits. They use the phenomenon of polarization to
derive $N$ virtual channels. Information is then transmitted over $K$ reliable
channels listed in the information set $\mathcal{I}$, while the remaining (unreliable)
channels are set to zero and called frozen bits (frozen set $\mathcal{F}$)
\cite{ari_2009}. Throughout the paper, we use the following notation:
$\mathbf{x} \in \{0,1\}^N$ is the code word and $\hat{\mathbf{x}}$ its estimation after decoding; $\mathbf{y}$ represents the vector of received channel \glspl{LLR}.

\subsection{Successive Cancellation Decoding}

\SC decoding can be described as depth-first tree traversal of the \gls{PFT} \cite{alaksc_2011}.
The \PFT has $\log(N)+1$~stages~$s$ and $N$ leaf nodes at stage $s=0$,
representing frozen bits and information bits. Each node $v$ receives a
\gls{LLR}-vector $\alpha^v$ of size~$N_s$ to first calculate the elements
of the vector $\alpha^l$
\begin{equation}
    \alpha^l_i=f\left(\alpha^v_i, \alpha^v_{i+N_v/2}\right)
    \label{eq:a_l}
\end{equation}
by the hardware-efficient min-sum formulation of the f-function
\begin{equation}
    f(a,b) = \sign\left(a\right) \sign\left(b\right) \min\left(\left|a\right|, \left|b\right| \right).
	\label{eq:f_fun}
\end{equation}
The resulting vector $\alpha^l$ of size $N_s/2$ is passed to the left child. With
the bit vector $\beta^l$ received from the left child, the $N_s/2$ elements of
$\alpha^r$ are calculated by the g-function
\begin{equation}
    \alpha^r_i=g\left(\alpha^v_i,\alpha^v_{i+N_v/2}, \beta^l_i\right)
    \label{eq:a_r}
\end{equation}
with
\begin{equation}
    g(a,b,c)=\left(1-2c\right)\cdot a + b,
    \label{eq:g_fun}
\end{equation}
and sent to the right child. With the results of both children $\beta^l$ and
$\beta^r$, the partial sum $\beta^v$ is calculated and send to the parent node.
In the leaf nodes, the bit decisions are made. Frozen bits are, per definition,
always $0$, information bit nodes return
\begin{equation}
    \beta^v = \sign \left( a^v \right),
\end{equation}
which is $0$ if $a^v \ge 0$ and $1$ otherwise. The decoder output $\hat{\mathbf{x}}=\operatorname{SC}(y)$ is equal to the output $\beta^v$ of the root node \cite{alaksc_2011}.

\subsection{Fast-SSC Decoding}
\label{sec:fastssc}

The \PFT can be pruned to reduce the number of operations to decode one
code word. Sub-trees containing only frozen bits don't have to be traversed,
since their decoding result is known to be an all-zero vector in advance. Such
subtrees are replaced by \RFr nodes. Similarly, subtrees without any frozen
bits can be decoded directly by \gls{HDD}, since no parity information is
contained. Therefore, these nodes are called Rate-1 nodes \cite{alaksc_2011}.

Further optimizations were proposed by \cite{sargia_2014}, denoted Fast-
\gls{SSC} decoding. If a sub-tree contains only one information bit, it is
considered a \REP code and replaced by a specialized \REP node. All bits with
index $j$ of a \REP node are decoded according to:
\begin{equation}
    \beta^{v,\text{REP}}_j = \sign \left( \sum_{k=0}^{N_s-1} \alpha^v_k \right).
    \label{eq:rep}
\end{equation}

In subtrees containing only one frozen bit, this bit always acts as parity bit.
Thus, the partial sum of this sub-tree represents a \SPC code. A specialized
\SPC node performs \ML decoding by calculating the parity $ \gamma^v \in \{0,1\}$
of the input:
\begin{equation}
    \gamma^v = \bigoplus_{j=0}^{N_s-1}\sign \left( \alpha^v_j \right),
    \label{eq:parity_spc}
\end{equation}
finding the least reliable bit
\begin{equation}
    j_{\min} = \argmin_{j\in[0,N_s)}\left|\alpha^v_j\right|
    \label{eq:j_min_spc}
\end{equation}
and setting $\beta^v$ to satisfy the single parity constraint:
\begin{equation}
    \beta^{v,\text{SPC}}_j = \begin{cases}
        \sign\alpha^v_j\oplus\gamma^v &\text{if } j = j_{\min}\\
        \sign\alpha^v_j               &\text{otherwise.}
              \end{cases}
    \label{eq:spc}
\end{equation}

\section{Automorphism Ensemble Decoding}
\label{sec:aed}
\gls{AED} was proposed in \cite{geielk_2021} as a general method to improve the error-rate performance by running an ensemble of $M$ independent \gls{SC} decoders. Each decoder $m$ operates on a permuted version of the received sequence and each decoding result is then unpermuted as
\begin{equation}
    \hat{\mathbf{x}}_m = \pi_m^{-1}\left(\operatorname{SC}\left(\pi_m(\mathbf{y})\right)\right).
\end{equation}
The permutations $\pi_m$ stem from the automorphism group of the code, which is the set of permutations of the coded bits that map every code word onto a (not necessarily different) code word. Permuted received sequences therefore correspond to different noise realizations which may be easier to decode. Hence, by choosing the most likely candidate from $\hat{\mathbf{x}}_m$, the probability of finding the correct code word is increased.

\subsection{Automorphisms and Code Construction}
\label{sec:aed_code}
The automorphism group of polar codes contains the \gls{BLTA} group which is defined as
\begin{equation}
    \mathbf{z}' = \mathbf{A}\mathbf{z}+\mathbf{b},
\end{equation}
where $\mathbf{A}$ is an invertible, block lower triangular binary matrix,
$\mathbf{b}$ an arbitrary binary vector, and $\mathbf{z},\mathbf{z}'$ are the
binary representations of the bit indices before and after permutation,
respectively \cite{geielk_2021a}. The block profile of $\mathbf{A}$, and thus,
the number of automorphisms, is dependent on the choice of the information set
$\mathcal{I}$. Therefore, for good performance under \gls{AED}, the polar code
design is critical. In \cite{pilbio_2022}, the $\mathcal{P}(128,60)$ code with
minimum information set $\mathcal{I}_\mathrm{min}=\{27\}$ was found to be
particularly well suited for \gls{AED}, as its \gls{BLTA} group has block
profile $\mathbf{s}=(3,4)$.

\subsection{Automorphism Selection}\label{sec:aut_sel}
It was shown in \cite{pilbio_2022} that permutations can be partitioned into
equivalence classes. Within each \gls{EC}, a permuted decoding results in the
same code word estimate. Hence, at most one permutation from each \gls{EC}
should be used in \gls{AED}. As the  $\mathcal{P}(128,60)$ code exhibits 2205
\glspl{EC}, a subset of \glspl{EC} has to be selected for use in \gls{AED}.

Additionally, it is beneficial to have a single list of $M$ permutations and
smaller ensemble sizes $M'<M$ use a subset of this list. As there is no theory
yet on how to select a good subset of \glspl{EC}, we propose a greedy,
data-driven method that works as follows:

\begin{enumerate}
    \item Generate one permutation from each \gls{EC}.
    \item Due to symmetry, select an arbitrary permutation $\pi_1$.
    \item Collect a batch of noisy code words that cannot be correctly decoded with any of the selected permutations.
    \item Decode this batch with the permutations that have not been selected yet and select the one that can correctly decode most of the noisy code words within the batch.
    \item Repeat steps 3 to 5 until $M$ permutations $\pi_1,\dots,\pi_M$ have been selected.
\end{enumerate}

\subsection{Conventional Candidate Selection}
\label{sec:can_sel}

To select the most probable code word $\hat{\mathbf{x}}$ of all $M$ decoding results $\hat{\mathbf{x}}_m$,
\cite{geielk_2021} propose the ML-in-the-list method
\begin{equation}
    \hat{\mathbf{x}} = \argmax_{\hat{\mathbf{x}}_m, m \in \{1,...,M\}}\mathbf{y}^\mathrm{T} (1-2\hat{\mathbf{x}}_m)
    \label{eq:ml}
\end{equation}
that chooses the code word $\hat{\mathbf{x}}_m$ that correlates most with the received \glspl{LLR}.

\subsection{Path Metric}

A drawback of the ML-in-the-list method is that the received channel \LLRs
$\mathbf{y}$ are necessary to perform the final decision. This requires a large
buffer in high-throughput decoders that are deeply pipelined. This buffer can
account up to 25\% of the total decoder area, dependent on $M$. This memory is
also a major source of power consumption. Hence, it is important to find
techniques to avoid this buffer. We propose a new \PM based candidate selection
for \AED with \SC constituent decoders. This \PM technique replaces the
ML-in-the-list selection and, thus, increases the locality in the \AED
implementation. Moreover, it makes the buffer redundant.

A \PM is used in \SCL decoding to rate the different decoding paths. An
efficient \gls{LLR}-based \PM calculation was presented in \cite{balpar_2015}.
In particular, it was shown that a \PM calculated as
\begin{equation}
    \text{PM} = \sum_{i = 0}^{N-1} \ln \left(1+e^{-(1-2\beta^v_i)\alpha^v_i}\right) \label{eq:pm_exact}
\end{equation}
is a decreasing function with the likelihood of the code word candidates.
Therefore, the path with the lowest \PM corresponds to the most likely
transmitted code word. The terms in (\ref{eq:pm_exact}) have the
implementation-friendly approximation
\begin{equation}
    \ln (1+e^x) \approx \max(0,x).
\end{equation}

In contrast to \SCL decoding, \SC decoding does not split the decoding path.
For information bits ($i\in \mathcal{I}$), $\hat{u}_i$ is equal to the hard
decision on $\alpha^v_i$ and, therefore, they do not contribute to the PM.
Thus, for \SC decoding, it is sufficient to update the \PM only for frozen bits,
where only $\alpha^v_i < 0$ indicates a transmission error.
This leads to the following cost function for the \PM in the leaf nodes:
\begin{equation}
    \text{PM} = \sum_{i \in \mathcal{F}}
        \lvert \min(0,\alpha^v_i) \rvert
    \label{eq:pm_fb_sum}
\end{equation}

The code word $\hat{\mathbf{x}}_m$ with the smallest \PM is then selected as
output of \gls{AED}.

\subsection{Fast-SSC Path Metric}
\label{sec:fastssc_pm}

State-of-the-art high-throughput low-latency decoder architectures are based on
the Fast-SSC decoding algorithm \cite{giabal_2017}. Since Fast-SSC decoding
operates on higher stages in the \gls{PFT}, the \LLR values of the frozen leaf
nodes are not calculated. Therefore, the \PM calculation presented previously
is not directly applicable. Thus, we derive \PM calculation methods for the
specialized nodes analogous to Fast \gls{SSCL} decoding~\cite{hascon_2017}. The value
of $\text{PM}^{(\ell)}$ is accumulated sequentially, where $\ell$ denotes the
index of the leaf node in the pruned \gls{PFT} and is initialized with
$\text{PM}^{(0)}=0$.

As discussed above, the \SC decoder does not split the decoding path and the \PM
calculation method is only needed for specialized nodes containing frozen bits,
\ie \gls{R0}, \REP and \SPC nodes:

\subsubsection{\glsfirst{R0} nodes}

For the \RFr node, we follow (\ref{eq:pm_fb_sum}), but execute the calculation
directly on the input \LLR vector $\alpha^v$ of the node with node size $N_s$:

\begin{equation}
    \text{PM}^{(\ell)} =\text{PM}^{(\ell-1)} +
        \sum_{j=0}^{N_s-1} \lvert \min(0,\alpha^v_j) \rvert.
    \label{eq:pm_r0_sum}
\end{equation}

\subsubsection{\glsfirst{REP} nodes}

According to (\ref{eq:rep}), the repeated bit decision is based on the sign of
the sum of all \LLRs related to the \REP node. Thus, corrected bits are located
at the positions $j$, for which $\sign(\alpha^v_j)\neq\beta^v_j$. The \PM
consequently is calculated by

\begin{equation}
    \text{PM}^{(\ell)} =\text{PM}^{(\ell-1)} +
        \sum_{\substack{j=0 \\ \sign(\alpha^v_j)\neq\beta^v_j}}^{N_s-1}
        \lvert \alpha^v_j \rvert.
    \label{eq:pm_rep_sum}
\end{equation}

\subsubsection{\glsfirst{SPC} nodes}

For the \SPC node, the decoding includes the correction of the least reliable
bit to satisfy the single parity constraint
((\ref{eq:parity_spc})-(\ref{eq:spc})). Thus, the \PM can be derived as:

\begin{equation}
    \text{PM}^{(\ell)} =\text{PM}^{(\ell-1)} +
        \gamma^v\cdot\lvert \alpha^v_{j_{\min}} \rvert.
    \label{eq:pm_spc}
\end{equation}

Simulations verified that this Fast-SSC \PM results in identical error correction performance as the conventional candidate selection based on (\ref{eq:ml}).

\subsection{Architecture}

\begin{figure}
	\includegraphics[width=1\linewidth]{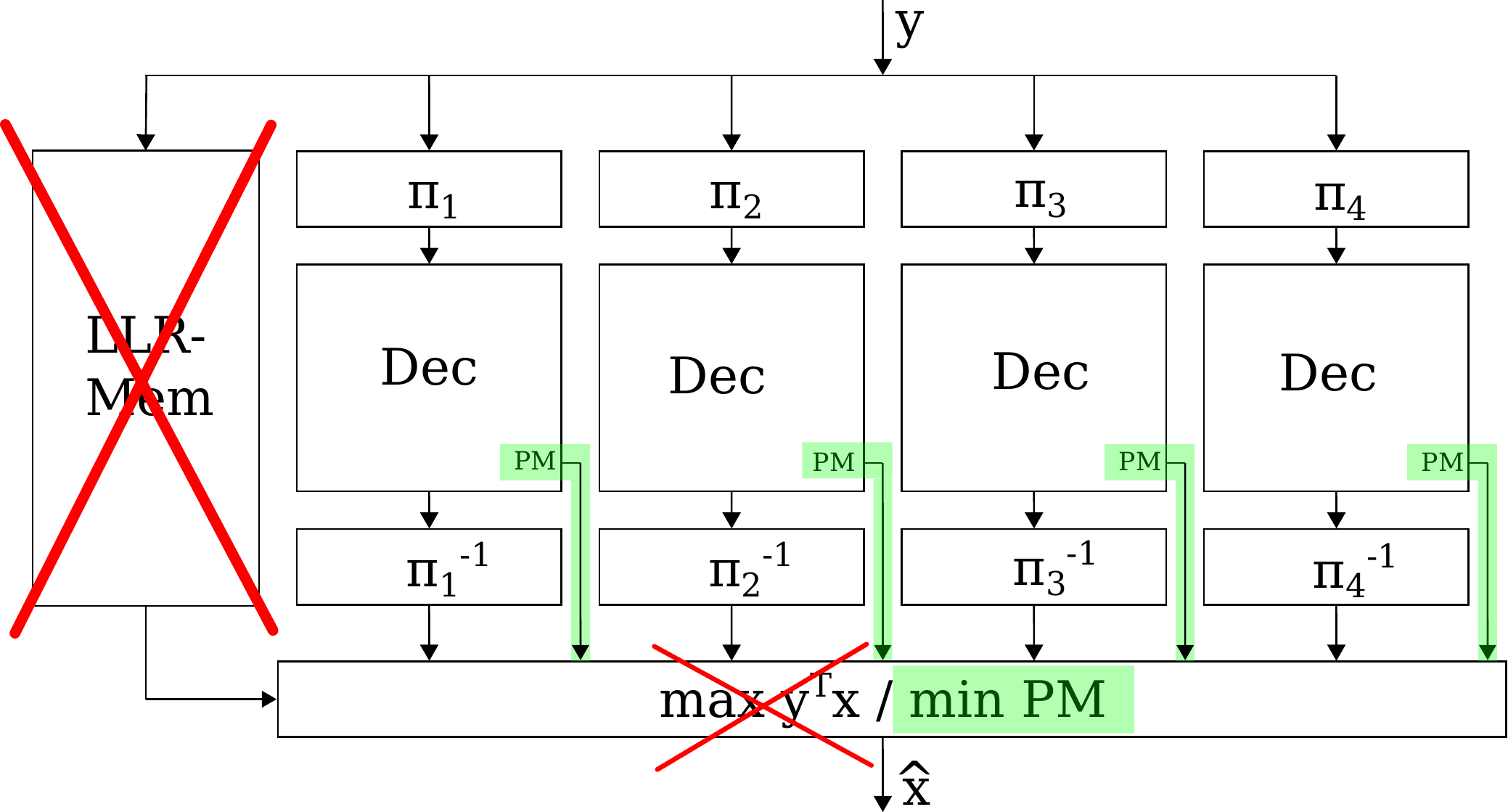}
	\centering
    \caption{Path metric based \AED architecture with $M=4$. In contrast to
    conventional \gls{AED}, the proposed architecture does not require a memory
for the \glspl{LLR}. The novel path metric based selection is highlighted in
green.}
    \label{fig:architecture}
\end{figure}

Figure \ref{fig:architecture} shows an \AED architecture with $M=4$ that
consists of $M$ identical \SC decoders. The permutations $\pi_m$ and
$\pi_m^{-1}$ map the \glspl{LLR} $\mathbf{y}$ and code words related to each SC
decoder instance, respectively, according to the automorphisms
(Section~\ref{sec:aut_sel}). This mapping does not require any logic or
registers. All decoders work independently and output their
decoding results as well as the PM. This \PM is calculated in each decoder as
described in Section \ref{sec:fastssc_pm} and fed into the selection unit, which
then outputs the code word estimate $\hat{x}$ with the lowest \PM value. All
decoders are generated automatically by a framework as fully unrolled and
pipelined architectures.

To illustrate the proposed optimization, Figure \ref{fig:architecture} also
shows the code word selection of the conventional \AED architecture with a
memory for \glspl{LLR} and the ML-in-the-list method but crossed out in red, while the novel path metric based selection is highlighted in green.

\section{Results}
\label{sec:results}

To enable a fair comparison, all decoders (\AED and \SCL decoders) were
optimized for the same throughput, \ie \Gbps{64}. We investigated different
degrees of parallelism $M\in\{2,4,8,16\}$ for the AE decoders and list sizes
$L\in\{2,4,8,16\}$ for the \SCL decoders. These decoders are denoted
\gls{AED}-$M$ and \gls{SCL}-$L$, respectively.  The \SCL decoders use the
optimizations from \cite{johkes_2022} and are, to the best of our knowledge, the
most efficient \SCL decoder implementations for very high throughput.

While this high throughput might not be necessary for the \gls{URLLC} use case, the
investigated unrolled and pipelined architectures are not only highly area and energy efficient,
but also offer very low latency, which is one of the core requirements for \gls{URLLC}.
With a reduction in frequency, the throughput could be decreased to reduce power consumption, but at the cost of increased latency.

All designs were synthesized with \textit{Design Compiler} and placed and
routed with \textit{IC-Compiler}, both from \textit{Synopsys} in a \nm{12}
FinFET technology from Global Foundries under worst case \gls{PVT} conditions
(\SI{125}{\degreeCelsius}, \SI{0.9}{V} for timing, \SI{1.0}{V} for power).

Error correction performance was simulated for an \gls{AWGN} channel and
\gls{BPSK} modulation. A minimum of 100 erroneous code words were simulated.
To assess the error correction capability of the codes, we also evaluated the
ML performance. For this, we simulated two bounds: an upper ML bound and a lower
ML bound. If both bounds match, we get ML performance. The upper ML bound was
calculated by performing \SCL decoding with $L=128$ and counting the cases
where a code word is closer to the received channel values than the correct
code word. The lower bound is derived in the same way, but additionally the
correct code word has to be in the list of candidates.

\subsection{Comparison with SCL}
\label{subsec:aed_scl}

\begin{figure}
	\includegraphics[width=1\linewidth]{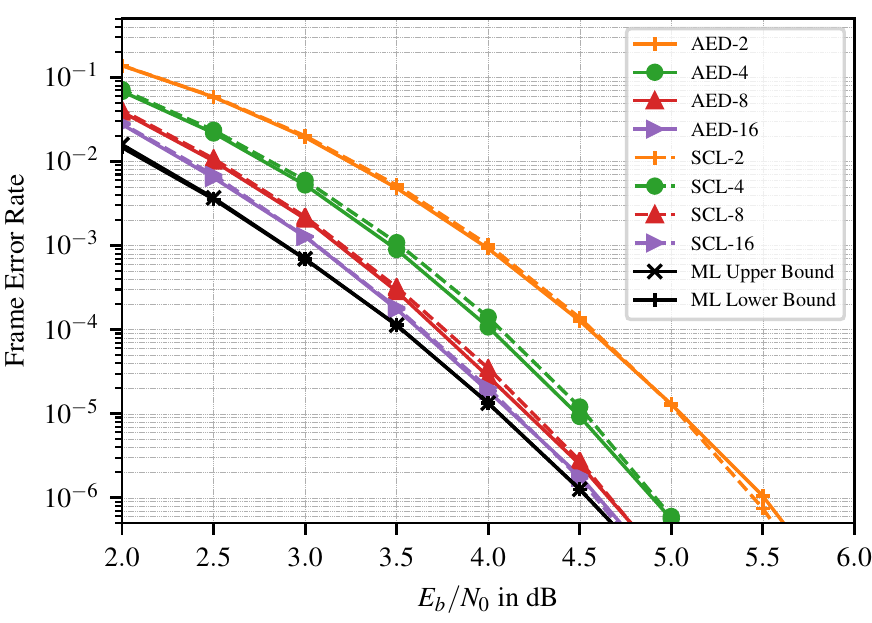}
	\centering
    \caption{AED vs. SCL decoding for the same $\mathcal{P}(128,60)$}
    \label{fig:plot_aed_scl}
\end{figure}

We first compare the error correction performance of \AED with \SCL decoding
(Figure \ref{fig:plot_aed_scl}) using the same $\mathcal{P}(128,60)$ as
described in Section~\ref{sec:aed_code}. Both decoding algorithms perform very
similar over the complete range of $L$ / $M$. They both benefit from a high minimum
distance. We therefore omitted an additional \CRC since it doesn't offer any
improvement over the chosen code. For $L=M=16$ both decoding algorithms
approach the ML Bounds.

Table~\ref{tab:aed_scl} shows the implementation results. We observe that \AED
outperforms its \SCL counterpart in all presented metrics. With increasing
$L$ / $M$, the gap between \AED and \SCL expands, since the \AED implementations
largely benefit from locality. For $L=M=16$, a
more than 8.9$\times$ better area efficiency, a 4.5$\times$ better energy
efficiency and a 5.3$\times$ shorter latency are observed.

\begin{table*}[ht]
    \centering
    \caption{
        Implementation results of AED and SCL decoder for the same
        $\mathcal{P}(128,60)$
    }
    \label{tab:aed_scl}
    \begin{tabular}{lcccccccc}
\toprule
 &AED-2&SCL-2&AED-4&SCL-4&AED-8&SCL-8&AED-16&SCL-16 \\
\midrule
Frequency {[}MHz{]}       &498 &476 &497 &462 &498 &478 &495 &403 \\
Throughput {[}Gbps{]}     &63.7 &61.0 &63.6 &59.1 &63.7 &61.2 &63.3 &51.6 \\
Latency {[}CC{]}          &11 &21 &11 &27 &11 &39 &11 &48 \\
\textbf{Latency {[}ns{]}} &\textbf{22.1 }&\textbf{44.1 }&\textbf{22.1 }&\textbf{58.5 }&\textbf{22.1 }&\textbf{81.5 }&\textbf{22.2 }&\textbf{119.1 }\\
Area {[}mm$^2${]}         &0.045 &0.064 &0.087 &0.154 &0.170 &0.450 &0.338 &2.464 \\
\textbf{Area Eff. {[}Gbps/mm$^2${]}}&\textbf{1407.9 }&\textbf{956.0 }&\textbf{733.8 }&\textbf{385.0 }&\textbf{375.1 }&\textbf{136.2 }&\textbf{187.0 }&\textbf{20.9 }\\
Utilization {[}\%{]}      &73 &73 &73 &72 &74 &74 &73 &47 \\
Power Total {[}W{]}       &0.07 &0.09 &0.15 &0.22 &0.32 &0.61 &0.64 &2.38 \\
\textbf{Energy Eff. {[}pJ/bit{]}}&\textbf{1.16 }&\textbf{1.48 }&\textbf{2.32 }&\textbf{3.73 }&\textbf{5.01 }&\textbf{10.02 }&\textbf{10.12 }&\textbf{46.21 }\\
Power Density {[}W/mm$^2${]}&1.63 &1.42 &1.70 &1.44 &1.88 &1.36 &1.89 &0.97 \\
\bottomrule
\end{tabular}
\end{table*}

\subsection{Comparison with 5G}

\begin{figure}
	\includegraphics[width=1\linewidth]{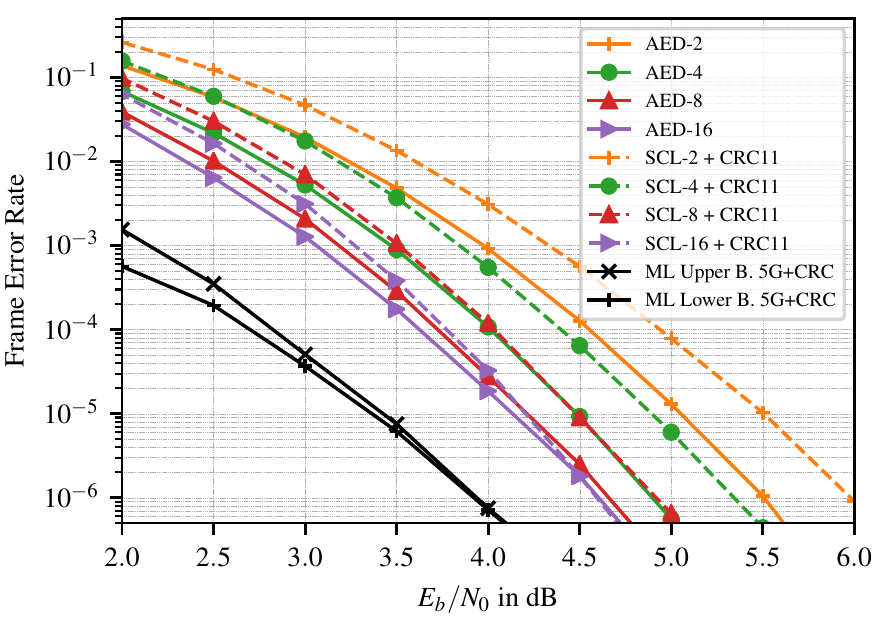}
	\centering
    \caption{AED for $\mathcal{P}(128,60)$ vs. CRC-aided SCL decoding for 5G
        $\mathcal{P}(128,60+C)$}
    \label{fig:plot_aed_5g}
\end{figure}

We also compare our implementations with optimized state-of-the-art CRC-aided
\SCL decoders for a 5G $\mathcal{P}(128,60+C)$ with an \CRC length of $C=11$. As
shown in Figure~\ref{fig:plot_aed_5g}, AED outperforms CRC-aided \SCL decoding
for all $L$ / $M$, while for $M=L=16$ the difference is shrinking for higher
\gls{SNR}. Note that AED-4 provides very similar performance as the CRC-aided
SCL-8 and would benefit from additional hardware savings under this
error correction requirement. While the ML Bounds for the 5G
$\mathcal{P}(128,60+C)$ are lower than for the $\mathcal{P}(128,60)$ code, the
\SCL decoders cannot benefit from that advantage.

Analyzing the implementation costs of the \AED and 5G \SCL decoders (Table~\ref{tab:aed_5g}),
implementations show a similar picture as in Section~\ref{subsec:aed_scl}.
However, the decoders for the 5G code show a slightly better area and energy
efficiency than for the $\mathcal{P}(128,60)$ code without CRC. This is caused
by the different code structure leading to different building blocks from the
\PFT traversal.

It is worth mentioning that the 5G code is optimized for a wide range of code
rates and usage scenarios, and not just for error correction performance and
implementation efficiency.

\begin{table*}[ht]
    \centering
    \caption{
        Implementation results of AED for $\mathcal{P}(128,60)$ and
        CRC-aided SCL decoders for 5G $\mathcal{P}(128,60+C)$
    }
    \label{tab:aed_5g}
    \begin{tabular}{lcccccccc}
\toprule
 &AED-2&SCL-2&AED-4&SCL-4&AED-8&SCL-8&AED-16&SCL-16 \\
\midrule
Frequency {[}MHz{]}       &498 &478 &497 &475 &498 &478 &495 &412 \\
Throughput {[}Gbps{]}     &63.7 &61.2 &63.6 &60.8 &63.7 &61.1 &63.3 &52.8 \\
Latency {[}CC{]}          &11 &21 &11 &26 &11 &36 &11 &47 \\
\textbf{Latency {[}ns{]}} &\textbf{22.1 }&\textbf{43.9 }&\textbf{22.1 }&\textbf{54.7 }&\textbf{22.1 }&\textbf{75.4 }&\textbf{22.2 }&\textbf{114.0 }\\
Area {[}mm$^2${]}         &0.045 &0.063 &0.087 &0.138 &0.170 &0.378 &0.338 &1.990 \\
\textbf{Area Eff. {[}Gbps/mm$^2${]}}&\textbf{1407.9 }&\textbf{968.8 }&\textbf{733.8 }&\textbf{440.0 }&\textbf{375.1 }&\textbf{161.8 }&\textbf{187.0 }&\textbf{26.5 }\\
Utilization {[}\%{]}      &73 &72 &73 &73 &74 &77 &73 &47 \\
Power Total {[}W{]}       &0.07 &0.08 &0.15 &0.21 &0.32 &0.62 &0.64 &1.84 \\
\textbf{Energy Eff. {[}pJ/bit{]}}&\textbf{1.16 }&\textbf{1.39 }&\textbf{2.32 }&\textbf{3.52 }&\textbf{5.01 }&\textbf{10.08 }&\textbf{10.12 }&\textbf{34.82 }\\
Power Density {[}W/mm$^2${]}&1.63 &1.35 &1.70 &1.55 &1.88 &1.63 &1.89 &0.92 \\
\bottomrule
\end{tabular}
\end{table*}

\subsection{Comparison with GRAND}

A comparison with the \gls{GRAND} or \gls{ORBGRAND} algorithm is difficult
since there exist not many implementations. In \cite{condo_2022} authors
present an optimized architecture in a \nm{7} technology for a
$\mathcal{P}(128,105)$. Its minimum latency, measured in clock cycles, is still
more than $2 \x$ the latency of our proposed AE decoder. Further comparison is
very difficult since the code differs and the authors provide synthesis results
only, while we performed full placement and routing under worst case \gls{PVT}
conditions.

\section{Conclusion}
\label{sec:conclusion}

In this paper, we presented a first \AED implementation and comparison with
state-of-the-art \SCL decoders. We show that \AED decoders outperform \SCL
decoders in latency, area and energy efficiency. We also demonstrate that the
error correction performance is comparable with the \SCL decoders and, in the
case of the 5G code, even exceeds the concatenation with CRC. \AED and the
proposed architecture is, thus, perfectly suited for the 6G \gls{URLLC} use
case, where low latency and high error correction is required.

\bibliographystyle{IEEEtran}
\bibliography{main}

\end{document}